\documentclass[aps,prd,preprint,floats,epsf,superscriptaddress,nofootinbib,amsmath]{revtex4}
\usepackage{amsmath,amssymb}
\usepackage{graphicx}
\usepackage{multirow}
\newcommand{\specialcell}[2][c]{%
  \begin{tabular}[#1]{@{}c@{}}#2\end{tabular}}

\begin{document}

{\small
\begin{flushright}
IPMU15-0073\\
\end{flushright} }

\bigskip

\title{Standard Model Effective Field Theory: \\
Integrating out a Generic Scalar}

\author{Cheng-Wei Chiang}
\email[e-mail: ]{chengwei@ncu.edu.tw}
\affiliation{Center for Mathematics and Theoretical Physics and Department of Physics, National Central University, Taoyuan, Taiwan 32001, R.O.C.}
\affiliation{Institute of Physics, Academia Sinica, Taipei, Taiwan 11529, R.O.C.}
\affiliation{Kavli IPMU (WPI), UTIAS, The University of Tokyo, Kashiwa, 277-8583, Japan}

\author{Ran Huo}
\email[e-mail: ]{ran.huo@ipmu.jp}
\affiliation{Kavli IPMU (WPI), UTIAS, The University of Tokyo, Kashiwa, 277-8583, Japan}

\date{\today}

\begin{abstract}
We consider renormalisable models extended in the scalar sector by a generic scalar field in addition to the standard model Higgs boson field, and work out the effective theory for the latter in the decoupling limit.  We match the full theory onto the effective theory at tree and one-loop levels, and concentrate on dimension-6 operators of the Higgs and electroweak gauge fields induced from such matching.  The Wilson coefficients of these dimension-6 operators from tree-level matching are further improved by renormalisation group running.  For specific $SU(2)_L$ representations of the scalar field, some ``accidental'' couplings with the Higgs field are allowed and can lead to dimension-6 operators at tree and/or one-loop level.  Otherwise, two types of interaction terms are identified to have only one-loop contributions, for the Wilson coefficients of which we have obtained a general formula.  Using the obtained results, we analyse constraints from electroweak oblique parameters and the Higgs data on several phenomenological models.
\end{abstract}


\maketitle

\section{Introduction}

After its discovery, the 125-GeV Higgs boson has been studied and found to be consistent with the standard model (SM) expectation as we know at present.  This observation suggests that any new physics degrees of freedom that directly couple with the SM-like Higgs boson should reside at a sufficiently high mass scale or be very weakly coupled with the SM particles so that they do not affect its properties significantly.  To study Higgs physics in this case, as in the case of Fermi theory, it is useful and satisfactory to work with an effective field theory (EFT) with higher dimensional operators of the SM fields organised in inverse powers of the new physics scale $\Lambda$.  Using the EFT approach, we can learn about possible types of new interactions at low energies.  By accumulating sufficient clues, a complete model of the new physics may be constructed.

As a start, we assume that new physics does not violate known gauge and Lorentz symmetries in the SM so that the higher dimensional operators obtained by integrating out the heavy degrees of freedom also satisfy the same symmetries.  There is only one dimension-5 operator (for one family of fermions) consistent with this, {\it i.e.}, the Weinberg operator that gives rise to Majorana mass for neutrinos~\cite{Weinberg:1979sa}.  This operator violates the lepton number by two units.  In the case of dimension-6 operators, the original attempt to compile a complete basis~\cite{Buchmuller:1985jz} was later found to be redundant~\cite{Grzadkowski:2003tf,Fox:2007in,AguilarSaavedra:2008zc}, leaving 64 independent operators (also for one family of fermions)~\cite{Grzadkowski:2010es} with five of them violating either baryon or lepton number~\cite{Weinberg:1979sa,Wilczek:1979hc,Abbott:1980zj}.  For weakly interacting renormalizable gauge theories that are perturbatively decoupling, the dimension-6 operators can be classified into potentially tree-generated and loop-generated ones~\cite{Arzt:1994gp,Einhorn:2013kja}.  Note that it has been stressed with explicit examples that the classification of higher dimensional operators into tree and loop ones within the EFT can be ambiguous~\cite{Jenkins:2013fya}.  A good discussion and comparison of different operator bases of popular choices~\cite{Buchmuller:1985jz,Hagiwara:1993ck,Giudice:2007fh} can be found in Ref.~\cite{Willenbrock:2014bja}.

There are some attractive motivations to consider models with an extended scalar sector.  For example, new scalar bosons in these models may facilitate a strong first-order phase transition for successful electroweak baryogenesis, provide Majorana mass for neutrinos, and/or have a connection with a hidden sector that houses dark matter candidates.  Even though it may not be possible to directly probe this sector due to the heavy masses of new scalar bosons and/or their feeble interactions with SM particles, they can nevertheless leave imprints in some electroweak precision observables.

In this paper, we analyse the EFT of the SM-like Higgs boson for a wide class of weakly coupled renormalisable new physics models extended by one type of scalar field(s)~\footnote{Multiple new scalar fields are allowed provided they have a common mass scale, as will be seen in the Zee-Babu model and the Georgi-Machacek model analysed in Sections~\ref{sec:ex}-A and \ref{sec:ex}-C.} and respecting CP symmetry.  It can be shown that only a few types of interactions contribute to the Higgs dimension-6 operators.  Two of them are $\mu H^\dagger H S$ and $\mu \left( H^\dagger\tau^a H \right) T^a$, where $\mu$ is a dimensionful quantity, $H$ is the SM Higgs doublet, $S$ is a singlet field, $T^a$ form a triplet field, and $\tau^a$ are the SU(2) generators.  These interactions only arise for specific representations of the new scalar and lead to dimension-6 operators when matching onto the EFT at tree level.  Another two are $(H^\dagger H)(\Phi^\dagger\Phi)$ and $(H^\dagger \tau^a H) (\Phi^\dagger t^a \Phi)$, where $t^a$ are the SU(2) operators appropriate for the new scalar field $\Phi$.  These interactions are more generic and, after the heavy scalar fields are integrated out, give rise to dimension-6 operators at one-loop level.  We work out the effective operators and the associated Wilson coefficients for an arbitrary new scalar field $\Phi (m,n,Y)$, where $m$ and $n$ denote its multiplicities under $SU(3)_C$ and $SU(2)_L$, respectively, and $Y$ is its hypercharge.  Phenomenological results of a few benchmark models are studied in this framework.

This paper is organised as follows.  In Section~\ref{sec:eff}, we define our framework of UV-complete models whose scalar sector is augmented from the SM by one new scalar field, and list dimension-6 operators composed of the SM Higgs and electroweak gauge fields that are of interest to us.  For the new scalar field of a generic representation, the dimension-6 operators are induced only from one-loop matching due to two specific types of quartic interactions in the UV theory.  We also identify accidental interaction terms for specific representations of the new scalar field that can lead to the dimension-6 operators already from tree-level matching.  We first concentrate on the accidental interactions in Section~\ref{sec:tree}, and work out the Wilson coefficients of induced operators from tree-level matching for these specific scalar representations.  These Wilson coefficients are further improved by renormalisation group running.  Section~\ref{sec:loop} discusses the matching of the full theory onto the effective theory at one-loop level for the new scalar of a general representation in the SM gauge group.  In Section~\ref{sec:ex}, we work out the results for a few benchmark models commonly considered in the literature.  Using the results, we show numerically how the model parameters are constrained by current and future electroweak precision observables and SM Higgs data.  Section~\ref{sec:sum} summarises our findings.

\section{Effective Operators and Wilson Coefficients \label{sec:eff}}

In the following, we will consider the renormalizable model having a scalar sector extended with a generic scalar field that couples to the SM Higgs field, and match it at tree and one-loop levels onto an effective theory with operators up to dimension-6.  Moreover, we will use the renormalization group equations (RGE's) to evolve the Wilson coefficients obtained from the tree-level matching from the new physics scale down to the electroweak scale, thereby capturing the leading-log loop corrections to them, and combine with those from the direct one-loop matching.  Throughout this paper, we will use $H$ and $\Phi$ to denote the SM Higgs field and the generic new scalar field, respectively.

\begin{table}[th]
\begin{tabular}{ccccc}
\hline\hline
~~Symbol~~ & Operator expression &~~~~~& ~~Symbol~~ & Operator expression \\
\hline
${\cal O}_6$ & $|H|^6$ &&
${\cal O}_{WW}$ & $g^2H^\dag HW^a_{\mu\nu}W^{a\mu\nu}$
\\
${\cal O}_H$ & $\frac12 \left( \partial_\mu |H|^2 \right)^2$ &&
${\cal O}_{BB}$ & $g'^2H^\dag HB_{\mu\nu}B^{\mu\nu}$
\\
${\cal O}_T$ & $\frac12 \left( H^\dagger \overleftrightarrow{D}_\mu H \right)^2$ &&
${\cal O}_{WB}$ & $2 g g' \left( H^\dagger \tau^a H \right) \left( W_{\mu\nu}^a B^{\mu\nu} \right)$
\\
${\cal O}_R$ & $ |H|^2\left( D_\mu H^\dagger D^\mu H \right)$ &&
${\cal O}_{W}$ & $i g \left( H^\dagger \overleftrightarrow{D}_\mu \tau^a H \right) D_\nu W^{a\mu\nu}$
\\
${\cal O}_{GG}$ & $g_s^2H^\dag HG^a_{\mu\nu}G^{a\mu\nu}$ &&
${\cal O}_{B}$ & $i g' \left( H^\dagger \overleftrightarrow{D}_\mu H \right) \partial_\nu B^{\mu\nu}$
\\
\hline\hline
\end{tabular}
\caption{Independent CP-even dimension-6 operators composed of only the Higgs and electroweak boson fields that are relevant to the analysis in this work.  Notations of fields and operators are explained in the main text.
\label{tab:Op}}
\end{table}

In Table~\ref{tab:Op}, we list ten CP-even dimension-6 operators composed of only the Higgs and electroweak gauge boson fields that are relevant for the electroweak precision and Higgs observables.  In the table, $D_\mu$ denotes the SM covariant derivative; $G^a_{\mu\nu}$, $W^a_{\mu\nu}$ and $B_{\mu\nu}$ are respectively the field strength tensors of the $SU(3)_C$, $SU(2)_L$ and $U(1)_Y$ groups with the associated gauge couplings $g_s$, $g$, and $g'$; and $A \overleftrightarrow{D}_\mu B \equiv A (D_\mu B) - (D_\mu A) B$.  The Wilson coefficient corresponding to the operator ${\cal O}_i$ will be denoted by $c_i$ and have mass dimension $-2$.

Assuming that the new scalar field $\Phi$ is a complex scalar, the kinetic and interaction terms relevant to our discussions are
\begin{equation}
\mathcal{L}\supset
(D_\mu\Phi)^\dag(D^\mu\Phi) - M^2\Phi^\dag\Phi
- \lambda(H^\dag H)(\Phi^\dag\Phi) - \lambda'(H^\dag\tau^aH)(\Phi^\dag t^a\Phi)
+ \mathcal{L}_\text{acc} ~,
\label{eq:lag}
\end{equation}
where $\mathcal{L}_\text{acc}$ denotes the ``accidental'' part to be detailed below.  For a real scalar field, terms quadratic in the new scalar should include an extra factor of $1/2$ and $\Phi^\dag$ is identified as $\Phi$.  In Eq.~(\ref{eq:lag}), $\tau^a = \sigma^a/2$ are the $SU(2)$ generators for the fundamental representation and $t^a$ are those for a generic representation, and the parameter $M$ sets the new physics scale that is assumed to be much higher than the electroweak scale.  Note that other terms such as the quartic interactions of the Higgs and the new scalar fields have been omitted, since they are irrelevant to the dimension-6 operators.  We will assume $\lambda,\lambda' > 0$ and the other quartic terms to be positive-definite as well to ensure that the potential is bounded from below.

The accidental part $\mathcal{L}_\text{acc}$ in Eq.~(\ref{eq:lag}) contains terms that are allowed only for specific representations of $\Phi$ and lead to dimension-6 operators from tree-level and/or one-loop matching.  It has two types of interactions.  The first one involves dimension-3 operators, and the only possibilities are:
\begin{equation}
\mathcal{L}_\text{acc,3}=\left\{
\begin{array}{ll}
- \mu H^\dag \Phi H  & \text{for a real singlet }(1,1,0),
\\
- \mu H^\dag \Phi^a t^a H  & \text{for a real triplet }(1,3,0),
\\
- \mu H \Phi^a t^a H + \mbox{h.c.}~ & \text{for a complex triplet }(1,3,-1),
\end{array}\right.
\label{eq:lag3}
\end{equation}
The second one involves dimension-4 operators that are only possible when $\Phi$ is an $SU(2)_L$ doublet.  For example,
\begin{equation}
\mathcal{L}_\text{acc,4}=
- \lambda'' (H^\dag\Phi)^2 - \lambda''' |H|^2 H^\dagger \Phi + \mbox{h.c.} \qquad \text{for a complex doublet }(1,2,1/2).
\label{eq:lag4}
\end{equation}
Note that the above term is allowed because $Y_\Phi = Y_H = 1/2$.  The effects of the $\lambda''$ terms have been discussed, for example, in Refs.~\cite{Henning:2014wua,Gorbahn:2015gxa}.  The $\lambda'''$ terms only lead to the operator ${\cal O}_6$ of no interest to our analysis.  New Yukawa terms involving $\Phi$ will also arise; yet they are irrelevant for our discussions.

\section{Tree-Level Matching and RGE Improvement \label{sec:tree}}

For the dimension-3 interaction terms in Eq.~(\ref{eq:lag3}), one can integrate out the new scalar field from the UV-complete theory by solving the associated equation of motion and plugging it back into the original Lagrangian~\footnote{A comprehensive study of the tree-level matching for general extensions of the SM with an arbitrary number and type of new scalar particles can be found in Ref.~\cite{deBlas:2014mba}}.  This gives rise to the following dimension-6 operators and some renormalization corrections for $|H|^4$ in the EFT:
\begin{equation}
\mathcal{L} \supset
\left\{
\begin{array}{ll}
\vspace{2mm}
\displaystyle
\frac{\mu^2}{M^4}{\cal O}_H  & \text{for a real singlet }(1,1,0),
\\
\vspace{2mm}
\displaystyle
\frac{\mu^2}{M^4}{\cal O}_T + 2\frac{\mu^2}{M^4}{\cal O}_R & \text{for a real triplet }(1,3,0),
\\
\displaystyle
\frac{\mu^2}{M^4}{\cal O}_H - \frac{\mu^2}{M^4}{\cal O}_T + 2\frac{\mu^2}{M^4}{\cal O}_R ~ &
\text{for a complex triplet }(1,3,-1).
\end{array}\right.
\label{eq:tree}
\end{equation}
Note that the operator ${\cal O}_T$ will lead to corrections of the oblique $T$ parameter, given by $T = \alpha_\text{EM}^{-1}c_Tv^2$ with the fine-structure constant $\alpha_\text{EM} = 1/128$ and $v\simeq 246$~GeV.  Therefore, the $T$ parameter measured to a high precision imposes a stringent constraint on the triplet models in Eq.~(\ref{eq:tree}), as the corresponding Wilson coefficients are not loop suppressed.  Using the measured electroweak $\rho$ parameter~\cite{PDG2014}, one can obtain an upper bound on $|\mu/M^2|$ to be $12.0 \times 10^{-5}$ and $3.4 \times 10^{-5}$~GeV$^{-1}$ at $95\%$ confidence level (CL) for the real and complex triplet cases, respectively.  These bounds can be translated into the corresponding bounds on the vacuum expectation values of the triplet field in the models.  By combining the real and complex triplet fields with a common $M$ parameter, corresponding to the Georgi-Machacek (GM) model, one gets a cancellation for the operator ${\cal O}_T$ so that $c_T = 0$.  That is, the GM model has only nonzero $c_H$ and $c_R$ at tree level.  We note in passing that the GM model also has other contributions from one-loop matching, which is to be discussed in Section~\ref{sec:ex}-C.

In addition to the above-mentioned contributions directly from tree-level matching, it is also possible to have additional corrections through RGE running of the other Wilson coefficients from the new physics scale $M$ to the electroweak scale, characterized by the $W$ boson mass $M_W$ (or sometimes the Higgs mass $M_h = 125$~GeV is used).  Here we will focus on the electroweak oblique corrections, of which $T$ has been discussed above and $S = 4\pi v^2 (4c_{WB}+c_W+c_B)$.  Note that with only tree-level matching and no RGE running, $S$ would be zero in the above-mentioned models.  The oblique $U$ parameter is not considered here because it first arises from a dimension-8 operator.  The anomalous dimensions for the RGE's are given in Ref.~\cite{Grojean:2013kd,Elias-Miro:2013mua,Jenkins:2013zja,Jenkins:2013wua,Alonso:2013hga,Elias-Miro:2013eta}.  It is well-known that the RGE anomalous dimensions have basis dependence.  Redundant operators would be radiatively generated even if one starts with an irreducible complete basis.  Using such a basis requires one to make use of equations of motion and/or field redefinitions to project the redundant Wilson coefficients generated in the running.  Since the redundant operator ${\cal O}_R$ already arises from tree-level matching in our analysis, we choose to work with a redundant basis containing ${\cal O}_R$ whose anomalous dimensions are given by~\cite{Elias-Miro:2013eta}:
\begin{align}
\frac{d}{d\ln Q}c_T &=
\frac{1}{(4\pi)^2}\frac{3}{2}{g'}^2 (c_H-c_R) + \cdots ~,
\nonumber
\\
\frac{d}{d\ln Q}c_W &=
-\frac{1}{(4\pi)^2} \frac{1}{3} (c_H-c_R) + \cdots ~,
\label{eq:RGE}
\\
\frac{d}{d\ln Q}c_B &=
-\frac{1}{(4\pi)^2} \frac{1}{3} (c_H-c_R) + \cdots ~,
\nonumber
\end{align}
where we have kept only the tree-level generated Wilson coefficients while leaving the other contributions in the ``$\cdots$'' parts.  Gauge independence of the running of these Wilson coefficients are checked and discussed at length in Ref.~\cite{Elias-Miro:2013eta}.

\begin{figure}[th]
\centering
\includegraphics[width=3in]{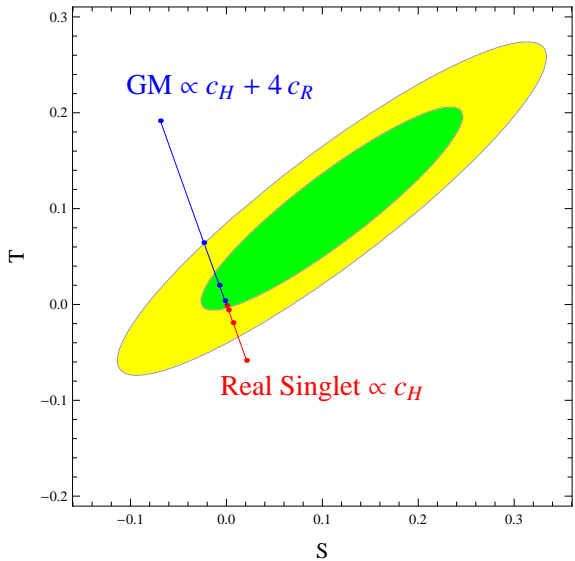}
\caption{Corrections to the $S$ and $T$ parameters from RGE-improved tree-level matching for the real singlet model (red line) and the GM model with $\lambda' = 0$ (blue line).  The inner green (outer yellow) ellipse corresponds to the $1\sigma$ ($2\sigma$) level.  The marks on each of the lines correspond to the parameter choices of $\mu=M=500$ GeV, 1 TeV, 2 TeV, and 5 TeV from the outer end toward the origin.}
\label{fig:treeRGE}
\end{figure}

It is readily seen that the list of models that can lead to dimension-6 operators from tree-level matching is rather limited~\cite{Henning:2014gca,Henning:2014wua}.  So is the list of models that can have oblique corrections induced at the tree-level matching with RGE improvements.  With the assumption of vanishing tree-level $T$ (but still allowing corrections from RGE), only two models are left: the real singlet model and the GM model.  In Fig.~\ref{fig:treeRGE}, we show the corrections to the $S$ and $T$ parameters from RGE-improved tree-level matching for the two models in red (the real singlet model) and blue (the GM model) lines.  Also shown in the plot are the measured ranges of the parameters at $1\sigma$ level in the inner green ellipse and $2\sigma$ level in the outer yellow ellipse.  In both cases, the oblique corrections arise solely from one-loop RGE running.  The slope of each line on the $S$-$T$ plane is fixed and determined by the tree-level Wilson coefficients $c_H$ and $c_R$.  It is $- 9 / (4\cos^2\theta_W)$ for both real singlet and GM models.  The exact location on each line is proportional to $\left( \mu^2v^2 / M^4 \right) \ln\left( M / M_W \right)$.  For each of the lines, we fix $\mu / M = 1$.  The marks from the outer end toward the origin have the parameter choices of $M=500$ GeV, 1 TeV, 2 TeV, and 5 TeV.  At the $2\sigma$ level, we find that the lower bound on the new scale is $M \agt 700$ GeV for the real singlet model and $M \agt 1000$ GeV for the GM model.  Since these bounds are obtained under the assumption that $\mu / M = 1$, they will thus become more relaxed (stringent) when $\mu < M$ ($\mu > M$).  Again, we note that the result for the GM model is purely based on the RGE-improved tree-level matching.  A more complete analysis including the one-loop matching will be presented in Section~\ref{sec:ex}-C.

\section{One-Loop Matching \label{sec:loop}}

Let's now turn our attention to the interactions that lead to dimension-6 operators only at the one-loop level.  Eq.~(\ref{eq:lag}) already contains all the relevant terms for $\Phi$ of a generic representation $(m,n,Y)$ under the SM gauge groups, where $m$ and $n$ denote its multiplicities under $SU(3)_C$ and $SU(2)_L$, respectively, and $Y$ is the hypercharge~\footnote{We use the normalisation that the electric charge of a particle $Q = I_3 + Y$ with $I_3$ being its third weak isospin component.}.  The argument goes as follows.  Any interaction term contributing in the covariant derivative expansion (CDE) approach to higher dimensional operators needs to be bilinear in the new scalar, and so is the SM Higgs fields according to our setup.  The SM Higgs bilinear under $SU(2)_L$ can only be decomposed as ${\bf 2} \otimes {\bf 2} \to {\bf 1} \oplus {\bf 3}$, with the singlet case corresponding to the $\lambda$ term and the triplet case to the $\lambda'$ term in Eq.~(\ref{eq:lag}).  If the new scalar field is a trivial representation of $SU(2)_L$, then only the $\lambda$ term is possible.  In other words, only a nontrivial representation ($n\geq2$) of $\Phi$ can give rise to the $\lambda'$ term.

After identifying the interaction terms, we then implement the CDE for the Coleman-Weinberg potential, as detailed in Refs.~\cite{Henning:2014gca,Henning:2014wua}.  Without going into details of the formalism, we just comment that what one needs are the coupling matrices of the new scalar field with the SM Higgs and gauge fields.  Since the coupling terms are all bilinear in the SM Higgs field and no linear term appears, the collection of contributing terms are limited and the calculations become straightforward.  For $\Phi$ of the generic representation $(m,n,Y)$, we find that the dimension-6 operators along with the associated Wilson coefficients induced by one-loop matching are given by
\begin{align}
\mathcal{L} \supset &
\frac{m}{(4\pi)^2} \left[
\frac{n\lambda^2}{6M^2}{\cal O}_H
+ \frac{n(n^2-1) {\lambda'}^2}{288 M^2}{\cal O}_T
+ \frac{n(n^2-1) {\lambda'}^2}{144 M^2}{\cal O}_R
\right.
\notag \\
& \qquad\qquad \left.
+ \frac{nY^2\lambda}{12M^2}{\cal O}_{BB}
+ \frac{n(n^2-1) \lambda}{144 M^2}{\cal O}_{WW}
+ \frac{n(n^2-1) Y\lambda'}{144 M^2}{\cal O}_{WB}
\right]
\notag \\
&
+ \frac{f(m)}{(4\pi)^2} \frac{n\lambda}{M^2} {\cal O}_{GG} ~,
\label{eq:general}
\end{align}
where the value of $f(m)$ depends on the representation $m$ of $SU(3)_C$.  For some simple representations, $f(1) = 0$, $f(3) = 1/24$, $f(6) = 5/24$ and $f(8) = 1/4$.  In Eq.~(\ref{eq:general}), we have assumed that the scalar is a complex field ($Y \not= 0$).  For a real scalar field, one should multiply an overall factor of $1/2$.  It is noted that each operator receives the contribution from either $\lambda$ or $\lambda'$, but not both.  The operators ${\cal O}_H$, ${\cal O}_{BB}$, ${\cal O}_{WW}$ and ${\cal O}_{GG}$ are induced by the $\lambda$ term, while ${\cal O}_T$, ${\cal O}_R$ and ${\cal O}_{WB}$ by the $\lambda'$ term and thus only for $\Phi$ of nontrivial $SU(2)_L$ representations.  The coefficients of ${\cal O}_H$, ${\cal O}_{BB}$ and ${\cal O}_{GG}$ are all linear in $n$, while those of ${\cal O}_{T}$, ${\cal O}_R$, ${\cal O}_{WW}$ and ${\cal O}_{WB}$ are proportional to the Dynkin index of the representation $\bf n$ and thus vanish for $SU(2)_L$ singlets.~\footnote{
In Ref.~\cite{Henning:2014wua}, it is pointed out that there are additional universal contributions to the pure gauge dimension-6 operators ${\cal O}_{2B}$, ${\cal O}_{2W}$, ${\cal O}_{3W}$, ${\cal O}_{2G}$ and ${\cal O}_{3G}$ defined in the reference.  The last two will not affect the electroweak and Higgs physics, while the first three are usually small in effect because they are proportional to the SM gauge couplings.  For completeness, we also quote the general result here and include them in the following fits:
\begin{equation}
\mathcal{L} \supset
\frac{m}{(4\pi)^2} \left[
\frac{nY^2g'^2}{30M^2}{\cal O}_{2B}
+ \frac{n(n^2-1)g^2}{360M^2}{\cal O}_{2W}
+ \frac{n(n^2-1)g^2}{360M^2}{\cal O}_{3W}
\right] ~.
\label{eq:hitoshi}
\end{equation}
}
We note in passing that tree-level perturbative unitarity of the $SU(2)_L$ interaction alone imposes a limit that the representation of a complex (real) scalar cannot be larger than an octet (nonet)~\cite{Hally:2012pu}.

\section{A Few Phenomenological Examples \label{sec:ex}}

\begin{table}[th]
\begin{tabular}{cccccc}
\hline\hline
Observable & ~~~~~~$\mu_{ZZ}$~~~~~~ & ~~~~~~$\mu_{WW}$~~~~~~ & ~~~~~~$\mu_{\gamma\gamma}$~~~~~~
& ~~~~~~$\mu_{bb}$~~~~~~ & ~~~~~~$\mu_{\tau\tau}$~~~~~~ \\
\hline
ATLAS\cite{ATLAS2015} & $1.44^{+0.40}_{-0.33}$ & $1.09^{+0.23}_{-0.21}$ & $1.17 \pm 0.27$
& $0.52 \pm 0.40$ & $1.43^{+0.43}_{-0.37}$ \\
CMS~\cite{Khachatryan:2014jba} & $1.00\pm0.29$ & $0.83\pm0.21$ & $1.12\pm0.24$ & $0.84\pm0.44$
& $0.91\pm0.28$ \\
\hline\hline
\end{tabular}
\caption{Signal strengths of various modes, indicated by the subscript in the first column, as measured at the LHC.}
\label{tab:sigs}
\end{table}

In this section, we explicitly work out the Wilson coefficients of relevant dimension-6 operators for a few well-motivated models whose scalar sector is extended with one new scalar field or multiple scalar fields of a common type, and discuss how the models are constrained by the electroweak precision data and the Higgs data, both for existing data and future expected measurements.  We have already seen that most such models have effective dimension-6 operators starting only at the one-loop level.  For current measurements we refer to the $U=0$ oblique parameter measurements of Ref.~\cite{Baak:2014ora} and the ATLAS (CMS) Higgs data with $4.5~(5.1)~\text{fb}^{-1}$ integrated luminosity at $\sqrt{s}=7$~TeV and $20.3~(19.7)~\text{fb}^{-1}$ integrated luminosity at $\sqrt{s}=8$~TeV~\cite{ATLAS2015,Khachatryan:2014jba}, as listed in Table~\ref{tab:sigs}.  We do not include tri-gauge boson precision measurements in our fitting.  For future expected sensitivities, we take the most aggressive oblique parameter measurements expected from the Tera Z experiment~\cite{TeraZ} and the projected Higgs data from Table~4 of Ref.~\cite{Gomez-Ceballos:2013zzn}.

\subsection{Zee-Babu model}

The Zee-Babu model~\cite{Zee:1985rj,Zee:1985id,Babu:1988ki} is one of the simplest phenomenological models that lead to dimension-6 operators only from the one-loop matching.  The model has one singly-charged and one doubly-charged singlet scalar fields without a color charge.  Assuming a common mass parameter $M$ for both of them and noticing that these fields cannot have the $\lambda'$ term in Eq.~(\ref{eq:lag}), we find that only the ${\cal O}_{BB}$ and ${\cal O}_H$ operators are induced at one-loop level:
\begin{equation}
\mathcal{L} \supset \frac{1}{(4\pi)^2}
\left[ \Big(\frac{\lambda_s}{12M^2} + \frac{\lambda_d}{3M^2}\Big){\cal O}_{BB}
+ \Big(\frac{\lambda_s^2}{6M^2} + \frac{\lambda_d^2}{6M^2}\Big){\cal O}_H
\right]~,
\end{equation}
where $\lambda_s$ and $\lambda_d$ are used to denote the coefficient $\lambda$ in Eq.~(\ref{eq:lag}) for the singly-charged and doubly-charged fields, respectively.
The operator ${\cal O}_{BB}$ results in a deviation in the Higgs diphoton decay rate from the SM expectation.  The ATLAS and CMS data can set a constraint on the Wilson coefficient: $\lambda_s / M^2 \lesssim 0.6~(1.2)\times10^{-4}~\text{GeV}^{-2}$ at $1\sigma~(2\sigma)$ level for the singly-charged scalar boson, and $\lambda_d / M^2 \lesssim 1.5~(3.0)\times10^{-5}~\text{GeV}^{-2}$ at $1\sigma~(2\sigma)$ level for the doubly-charged scalar boson.  The current bound for a doubly-charged scalar mass is about 400~GeV~\cite{ATLAS:2014kca} assuming $100$\% decay branching ratio to light leptons, while that for the singly-charged one is as low as 90~GeV \cite{Abbiendi:2003ji}.  Setting the new scalars at the corresponding mass bounds, the current measurements can constrain $\lambda_d\lesssim 2.3~(4.8)$ at $1\sigma$ ($2\sigma$) level for the doubly-charged scalar boson assuming $\lambda_s=0$, or $\lambda_s\lesssim 0.5~(1.0)$ at $1\sigma$ ($2\sigma$) level for the singly-charged scalar boson if $\lambda_d=0$.  Constraints on the quartic couplings of the model from perturbativity and stability can be found in Ref.~\cite{Herrero-Garcia:2014hfa}.

While current constraints mostly come from the Higgs diphoton rate, the universal Higgs correction ${\cal O}_H$, entering several channels to be well-measured at a lepton collider, can become equally or even more important in Higgs phenomenology.  The expected constraint for the singly-charged scalar boson at the future circular $e^+$-$e^-$ collider (FCC-ee) is $\lambda_s / M \lesssim 6\times10^{-3}~\text{GeV}^{-1}$ at $2\sigma$ level.  For the doubly-charged scalar boson, the constraint is still dominated by the Higgs diphoton rate, and $\lambda_d / M^2 \lesssim 6\times10^{-6}~\text{GeV}^{-2}$ at $2\sigma$ level is a good approximation for the mass regime of $M\lesssim800$~GeV.

\subsection{Two-Higgs doublet model}

As a second example for loop-induced dimension-6 operators, we consider the two-Higgs doublet model (2HDM).  We follow the notation and convention of Ref.~\cite{Haber:1993an}, except that each of our $\lambda$'s is twice bigger, and write down the scalar potential:
\begin{align}
\mathrm{V}_{\rm 2HDM} =&
m_{11}^2 (\Phi_1^{\dagger} \Phi_1) + m_{22}^2
(\Phi_2^{\dagger} \Phi_2) - m_{12}^2 (\Phi_1^{\dagger} \Phi_2) -
m_{12}^{*2}(\Phi_2^{\dagger} \Phi_1) \nonumber \\
&+ \lambda_1 (\Phi_1^{\dagger} \Phi_1)^2 +
\lambda_2 (\Phi_2^{\dagger} \Phi_2)^2 + 2\lambda_3 (\Phi_1^{\dagger}
\Phi_1)(\Phi_2^{\dagger} \Phi_2) + 2\lambda_4 (\Phi_1^{\dagger}
\Phi_2)(\Phi_2^{\dagger} \Phi_1)
\nonumber \\
&+ \lambda_5 (\Phi_1^{\dagger} \Phi_2)^2 +
\lambda_5^{*} (\Phi_2^{\dagger} \Phi_1)^2 ~,
\label{eq:V2HDM}
\end{align}
where we have left out the $\lambda_{6,7}$ terms that are forbidden by the $Z_2$ symmetry $\Phi_1 \to \Phi_1$ and $\Phi_2 \to - \Phi_2$.
As usual, we define the angle $\beta$ in terms of $\tan\beta = v_2 / v_1$, the ratio of the vacuum expectation values of $\Phi_2$ and $\Phi_1$, and the angle $\alpha$ as the rotation from the above basis to the mass eigenbasis for the CP-even neutral Higgs bosons.  In the decoupling limit, $\cos(\beta - \alpha) \to 0$ and the SM-like Higgs boson is much lighter than the other Higgs bosons.  In this limit, the above potential is turned into a form consistent with Eqs.~(\ref{eq:lag}) and (\ref{eq:lag4}) after the basis rotation of angle $\beta$.  We then determine
\begin{align}
\lambda =&
\frac{1}{8} \left[
3\lambda_1 + 3\lambda_2 + 10\lambda_3 + 2\lambda_4 - 6\lambda_5
- 3(\lambda_1 + \lambda_2 - 2\lambda_3 - 2\lambda_4 - 2\lambda_5) \cos4\beta
\right] ~,
\notag \\
\lambda' =&
\frac{1}{2} \left[
\lambda_1 + \lambda_2 - 2\lambda_3 + 6\lambda_4 - 2\lambda_5
- (\lambda_1 + \lambda_2 - 2\lambda_3 - 2\lambda_4 - 2\lambda_5) \cos4\beta
\right] ~.
\label{eq:lambda2HDM}
\end{align}

At this point, it is useful to make a comparison with results already existing in the literature, as we use a different parameterisation from others.  Using the identity $\tau^a_{ij}\tau^a_{kl}=\frac{1}{2}\delta_{il}\delta_{jk}-\frac{1}{4}\delta_{ij}\delta_{kl}$, 
we have $\lambda(H^\dag H)(\Phi^\dag\Phi) + \lambda'(H^\dag\tau^aH)(\Phi^\dag \tau^a\Phi)=(\lambda-\frac{1}{4}\lambda')(H^\dag H)(\Phi^\dag\Phi) + \frac{1}{2}\lambda'(H^\dag\Phi)(\Phi^\dag H)$.  We therefore make the identification $\lambda=\lambda_1+\frac{1}{2}\lambda_2$, $\lambda'=2\lambda_2$, where $\lambda_{1,2}$ here are those used in Ref.~\cite{Henning:2014wua}.  Without the $\lambda''$ terms in Eq.~(\ref{eq:lag4}), our most generic result corresponds to the ``Simpler 2HDM theory'' in Ref.~\cite{Henning:2014wua}.

\begin{figure}[th]
\centering
\includegraphics[height=3in]{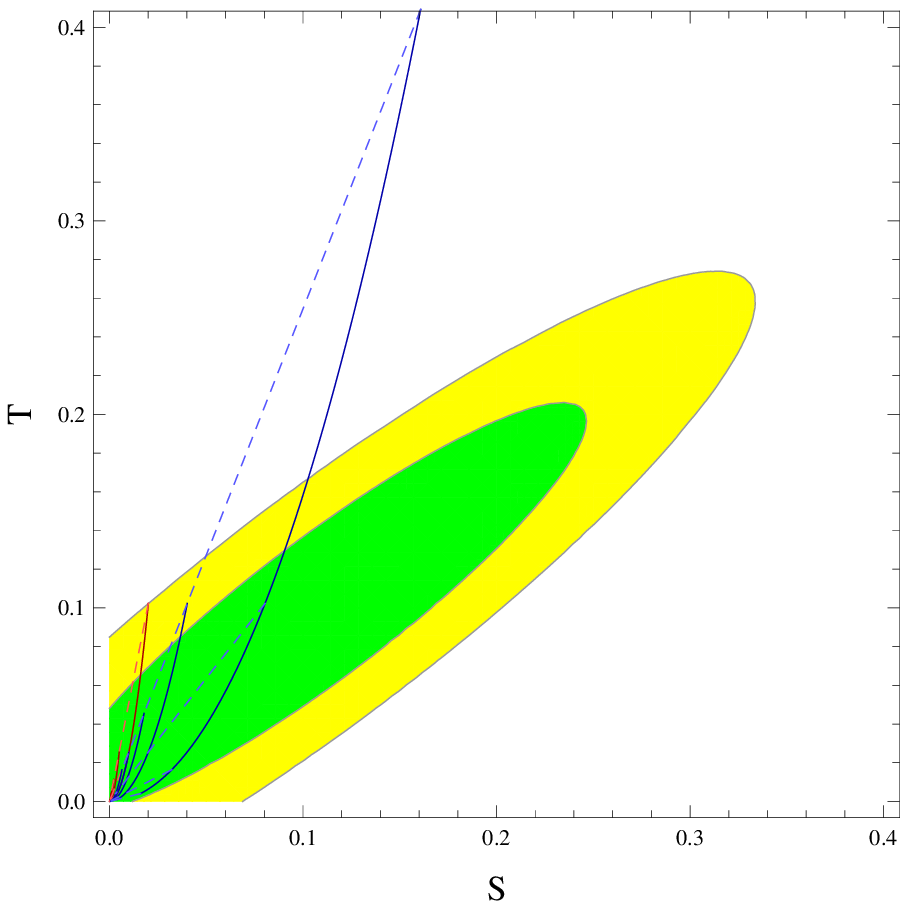}
\includegraphics[height=3in]{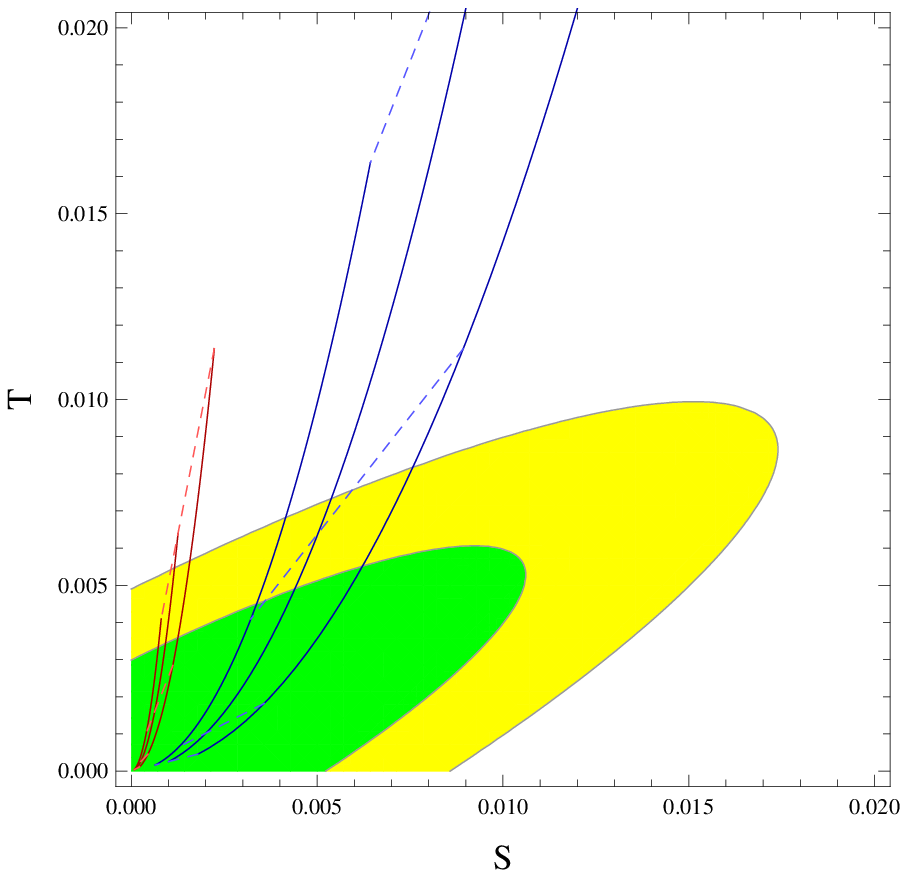}
\caption{Oblique corrections of the decoupling 2HDM (red), where we have omitted the accidental $\lambda''$ term in Eq.~(\ref{eq:lag4}), and $Y=1$ complex Higgs triplet model (blue), where we have omitted the RGE-improved tree-level contributions.  The left plot also shows the current LEP2 constraints at $1\sigma$ (green region) and $2\sigma$ (yellow region) levels, while the right plot shows the most aggressive constraints expected from the future Tera Z experiment.  For each set of the red and blue curves, the solid ones have fixed $M$ with the values of $200$~GeV, $400$~GeV, $600$~GeV from right to left in the left plot, and $600$~GeV, $800$~GeV, $1000$~GeV in the right plot.  The dashed lines have fixed couplings $\lambda'$ with the values of $2$, $1$, and $0.5$ from top to bottom.}
\label{fig:ST}
\end{figure}

In Fig.~\ref{fig:ST}, we use the red curves to show the behaviour of $S$ and $T$ parameters for the 2HDM in the decoupling limit.  For the set of solid red curves, we take $M = 200$, $400$ and $600$ GeV from right to left in the left plot and 600, 800, and 1000 GeV in the right plot.  It is seen that there is a sensitive dependence on the mass scale $M$.  From top to bottom, the dashed lines are for fixed $\lambda' = 2$, $1$, and $0.5$ in both plots.  One can see that the current precision of data barely constrains the case with $M = 200$ GeV and $\lambda' = 2$ at the $2\sigma$ level.  With sufficiently small $\lambda'$ ({\it e.g.}, $\alt 0.5$), even the future Tera Z experiment cannot probe the mass scale above about 200 GeV.

The Higgs sector of the minimal supersymmetric standard model (MSSM) is identical to the 2HDM except that the $\lambda$'s in Eq.~(\ref{eq:V2HDM}) should be replaced by either $\frac{1}{8}(g^2\pm g'^2)$ or $-\frac{1}{4}g^2$ at tree level.  Therefore, in the decoupling limit of the MSSM, the relevant interaction coefficients $\lambda$ and $\lambda'$ are $\mathcal{O}(0.5)$.  As seen in Fig.~\ref{fig:ST}, the oblique parameters are not very constraining even in the case of future lepton colliders.  In certain models such as the inert 2HDM where $\beta = 0$, $\lambda = 2\lambda_3 + \lambda_4$ and $\lambda' = 4\lambda_4$, which are not constrained by the SM Higgs coupling/mass and potentially large.  In this case, the precision measurements alone can probe a much larger parameter region.

\subsection{Georgi-Machacek model \label{sec:gm}}

Our third example is the GM model, which is special in that it is the only model that has contributions from both RGE-improved tree-level matching and the one-loop matching.  In addition to the SM Higgs doublet, the scalar sector of the GM model has a $Y=1$ complex triplet scalar field $X \equiv (\chi^{++},\chi^+,\chi^0/\sqrt{2})^T$ and a $Y=0$ real triplet scalar field $\Xi \equiv (\xi^+,\xi^0,\xi^-)^T$ with the scalar potential
\begin{align}
V_{\rm GM} =&
M^2 \left( X^\dagger X + \frac12 \Xi^\dagger \Xi \right)
+ \mu \left[ \left( H^\dagger \tau^a H^C \right) X^a + \mbox{h.c.} \right]
- \sqrt{2}\mu \left( H^\dagger \tau^a H \right) \Xi^a
\nonumber \\
&
+ \lambda (H^\dagger H) \left( X^\dagger X + \frac12 \Xi^\dagger \Xi \right)
+ \lambda' \left( H^\dagger \tau^a H \right) \left( X^\dagger t^a X \right)
\nonumber \\
&
+ \frac{\lambda'}{\sqrt{2}} \left( H^\dagger \tau^a H^C \right) \left( \Xi^\dagger t^a X + \mbox{h.c.}\right)
+ \cdots ~,
\label{eq:GM}
\end{align}
where $H^C = i \sigma^2 H^*$, terms have been written in accordance with Eqs.~(\ref{eq:lag}) and (\ref{eq:lag3}), and the ``$\cdots$'' part contains the irrelevant ones.  Note that the one-loop matching is not simply a sum of separate $Y=1$ complex and $Y=0$ real triplet contributions, as the last term mixes the real and complex fields.

Corrections to the oblique parameters are
\begin{align}
S &=
\frac{1}{4\pi} \left[
(-2) \frac{\mu^2v^2}{M^4} \ln\left(\frac{M}{M_W}\right) +\frac{2}{3} \frac{\lambda'v^2}{M^2}
\right]~,
\nonumber
\\
T &=
\frac{1}{4\pi} \left[
\frac{9}{2\cos^2\theta_W} \frac{\mu^2v^2}{M^4} \ln\left(\frac{M}{M_W}\right)
+ \frac{1}{96\pi \alpha_{\rm EM}} \frac{{\lambda'}^2v^2}{M^2}
\right]~.
\label{eq:GM}
\end{align}
Note that all these contributions are positive-definite.  The logarithmic part in each of the expressions comes from the one-loop RGE-induced Wilson coefficients after tree-level matching and is numerically presented in Fig.~\ref{fig:treeRGE}, while the second part is the result from one-loop matching.  To demonstrate purely one-loop contributions to the oblique parameters, we plot in Fig.~\ref{fig:ST} the blue curves for the case with a single $Y = 1$ complex triplet without the dimension-3 accidental interactions in Eq.~(\ref{eq:lag3}) using the same set of parameters $M$ and $\lambda'$ as for the 2HDM in the decoupling limit.  Apparently, the oblique parameter data are more constraining in this case.

\begin{figure}[t!h]
\centering
\includegraphics[height=3in]{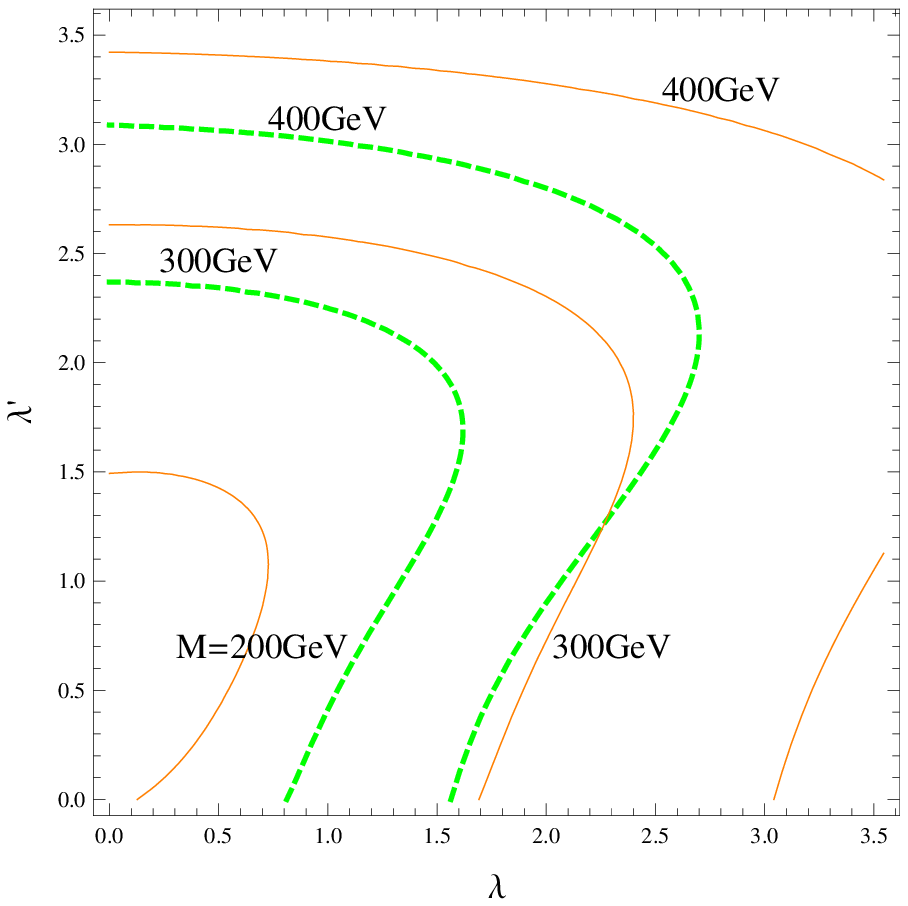}
\includegraphics[height=3in]{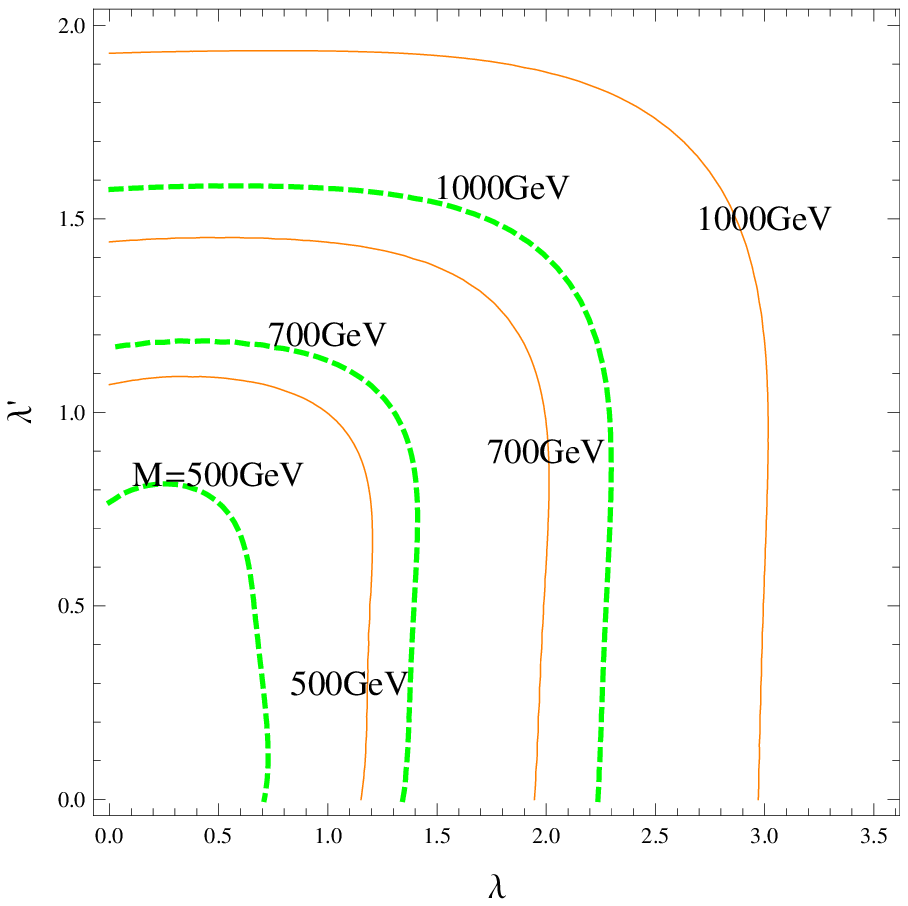}
\caption{Constraints on the $\lambda$-$\lambda'$ parameter space of the GM model from a global fit to both the electroweak oblique corrections and the Higgs data.  The left plot uses the current LEP2 oblique parameters and the current ATLAS and CMS Higgs data.  The right plot uses the most aggressive Tera Z result for the oblique parameters and the expected FCC-ee Higgs measurement constraints.  In the left (right) plot, we fix $\mu=50~(5)$~GeV.  Green dashed and orange curves are respectively contours at $1\sigma$ and $2\sigma$ levels for different choices of $M$.}
\label{fig:GM}
\end{figure}

By combining the RGE-improved tree-level and one-loop contributions to oblique parameters and the Higgs observables in Table~\ref{tab:sigs}, we perform a $\chi^2$ fit on the plane of $\lambda$ and $\lambda'$ for fixed values of $\mu$ and $M$.  Fig.~\ref{fig:GM} shows the $1\sigma$ and $2\sigma$ contours by fitting to the current data for $\mu=50$~GeV and $M = 200$, $300$ and $400$~GeV in the left plot, and by fitting to the future expectations for $\mu=5$~GeV and  $M = 500$, $700$ and $1000$~GeV in the right plot.  Note that in making these plots, we keep in mind that $\lambda^{(\prime)} \alt \sqrt{4\pi}$ to ensure the perturbativity.  If $\mu$ is chosen to have a larger (smaller) value, the allowed region will become more stringent (relaxed).

\subsection{Loop-generated neutrino mass models}

\begingroup
\squeezetable
\begin{table}[ht]
\centering
\renewcommand{\arraystretch}{1.2}
\begin{tabular}{c|c|c|c|c}
\hline\hline
Group & Model & SM Charge & Current constraints at $2\sigma$ level & Future constraints at $2\sigma$ level \\
\hline
\multirow{2}{*}{I} &
Zee & $(1,1,1)$ & $\frac{\lambda}{M^2}\lesssim1.2\times10^{-4}~\text{GeV}^{-2}$ & $\frac{\lambda}{M}\lesssim6\times10^{-3}~\text{GeV}^{-1}$ \\
&
Zee-Babu & $(1,1,2)$ & $\frac{\lambda}{M^2}\lesssim3\times10^{-5}~\text{GeV}^{-2}$ & $\frac{\lambda}{M^2}\lesssim6\times10^{-6}~\text{GeV}^{-2}$ \\
\hline
\multirow{5}{*}{II} &
\specialcell{$\bar{L^c}Q\phi$, $\bar{e^c}u\phi$ \\ $\bar{Q^c}Q\phi$, $\bar{u^c}d\phi$} & \specialcell{$(\bar{3},1,1/3)$ \\ $(3,1,-1/3)$} & $\frac{\lambda}{M^2}\lesssim7\times10^{-5}~\text{GeV}^{-2}$ & $\frac{\lambda}{M^2}\lesssim5\times10^{-6}~\text{GeV}^{-2}$ \\
&
$\bar{d^c}d\phi$ & $(3,1,2/3)$ & $\frac{\lambda}{M^2}\lesssim7\times10^{-5}~\text{GeV}^{-2}$ & $\frac{\lambda}{M^2}\lesssim5\times10^{-6}~\text{GeV}^{-2}$ \\
&
\specialcell{$\bar{e^c}d\phi$ \\ $\bar{u^c}u\phi$} & \specialcell{$(\bar{3},1,4/3)$ \\
$(3,1,-4/3)$} & $\frac{\lambda}{M^2}\lesssim3\times10^{-5}~\text{GeV}^{-2}$ & $\frac{\lambda}{M^2}\lesssim3\times10^{-6}~\text{GeV}^{-2}$ \\
\hline
\multirow{3}{*}{III} &
$\bar{Q^c}Q\phi$, $\bar{u^c}d\phi$ & $(\bar{6},1,-1/3)$ & $\frac{\lambda}{M^2}\lesssim1.3\times10^{-5}~\text{GeV}^{-2}$ & $\frac{\lambda}{M^2}\lesssim1\times10^{-6}~\text{GeV}^{-2}$ \\
&
$\bar{d^c}d\phi$ & $(\bar{6},1,2/3)$ & $\frac{\lambda}{M^2}\lesssim1.3\times10^{-5}~\text{GeV}^{-2}$ & $\frac{\lambda}{M^2}\lesssim1\times10^{-6}~\text{GeV}^{-2}$ \\
&
$\bar{u^c}u\phi$ & $(\bar{6},1,-4/3)$ & $\frac{\lambda}{M^2}\lesssim1.2\times10^{-5}~\text{GeV}^{-2}$ & $\frac{\lambda}{M^2}\lesssim9\times10^{-7}~\text{GeV}^{-2}$ \\
\hline
\multirow{2}{*}{IV} &
$\bar{d}L\phi$ & $(3,2,1/6)$ & $\frac{\lambda}{M^2}\lesssim3\times10^{-5}~\text{GeV}^{-2}$, $\frac{\lambda'}{M}\lesssim5\times10^{-3}~\text{GeV}^{-1}$ & $\frac{\lambda}{M^2}\lesssim2.5\times10^{-6}~\text{GeV}^{-2}$, $\frac{\lambda'}{M}\lesssim1.3\times10^{-3}~\text{GeV}^{-1}$ \\
&
$\bar{Q}e\phi$, $\bar{u}L\phi$ & $(3,2,7/6)$ & $\frac{\lambda}{M^2}\lesssim1.5\times10^{-5}~\text{GeV}^{-2}$, $\frac{\lambda'}{M}\lesssim6\times10^{-3}~\text{GeV}^{-1}$ & $\frac{\lambda}{M^2}\lesssim1.9\times10^{-6}~\text{GeV}^{-2}$, $\frac{\lambda'}{M}\lesssim1.6\times10^{-3}~\text{GeV}^{-1}$ \\
\hline
V &
$\bar{Q}u\phi$, $\bar{d}Q\phi$ & $(8,2,-1/2)$ & $\frac{\lambda}{M^2}\lesssim5\times10^{-6}~\text{GeV}^{-2}$, $\frac{\lambda'}{M}\lesssim3\times10^{-3}~\text{GeV}^{-1}$ & $\frac{\lambda}{M^2}\lesssim4\times10^{-7}~\text{GeV}^{-2}$, $\frac{\lambda'}{M}\lesssim7\times10^{-4}~\text{GeV}^{-1}$
\\
\hline
\multirow{2}{*}{VI} &
$\bar{L^c}Q\phi$ & $(\bar{3},3,1/3)$ & $\frac{\lambda}{M^2}\lesssim2\times10^{-5}~\text{GeV}^{-2}$, $\frac{\lambda'}{M}\lesssim3\times10^{-3}~\text{GeV}^{-1}$ & $\frac{\lambda}{M^2}\lesssim1.6\times10^{-6}~\text{GeV}^{-2}$, $\frac{\lambda'}{M}\lesssim7\times10^{-4}~\text{GeV}^{-1}$ \\
&
$\bar{Q^c}Q\phi$ & $(3,3,-1/3)$ & $\frac{\lambda}{M^2}\lesssim2\times10^{-5}~\text{GeV}^{-2}$, $\frac{\lambda'}{M}\lesssim2.5\times10^{-3}~\text{GeV}^{-1}$ & $\frac{\lambda}{M^2}\lesssim1.8\times10^{-6}~\text{GeV}^{-2}$, $\frac{\lambda'}{M}\lesssim6\times10^{-4}~\text{GeV}^{-1}$ \\
\hline
VII &
$\bar{Q^c}Q\phi$ & $(\bar{6},3,-1/3)$ & $\frac{\lambda}{M^2}\lesssim4\times10^{-6}~\text{GeV}^{-2}$, $\frac{\lambda'}{M}\lesssim2.5\times10^{-3}~\text{GeV}^{-1}$ & $\frac{\lambda}{M^2}\lesssim4\times10^{-7}~\text{GeV}^{-2}$, $\frac{\lambda'}{M}\lesssim5\times10^{-4}~\text{GeV}^{-1}$ \\
\hline\hline
\end{tabular}
\caption{More exotic scalar fields, their SM quantum numbers, current constraints and future constraints at the $2\sigma$ level.
\label{tab:neutrino}}
\end{table}
\endgroup

There are models with other types of exotic scalar fields.  Here we consider the exotic scalars introduced to induce effective $\Delta L=2$ operators at the loop level for generating Majorana mass for neutrinos.  Refs.~\cite{de Gouvea:2007xp,Angel:2012ug} provide a comprehensive list of such operators, which we reorganise into Table~\ref{tab:neutrino}.  In fact, the Zee-Babu model discussed in Section~\ref{sec:ex}-A belongs to this category (Group I in the table).  In the table, we also show estimates of the current constraints and the expected future constraints at the $2\sigma$ level for each type of exotic scalar field(s) by a fit using universally calculated Wilson coefficients.  These are bounds on $\lambda / M^2$ or sometimes $\lambda / M$ ($\lambda' / M$) when assuming $\lambda' = 0$ ($\lambda = 0$) and approximately valid for the regime of $M\leq800$ GeV.  It is noted that for individual models, more precise bounds require an EFT analysis at the full one-loop order.

\begin{figure}[t!]
\centering
\includegraphics[height=2.5in]{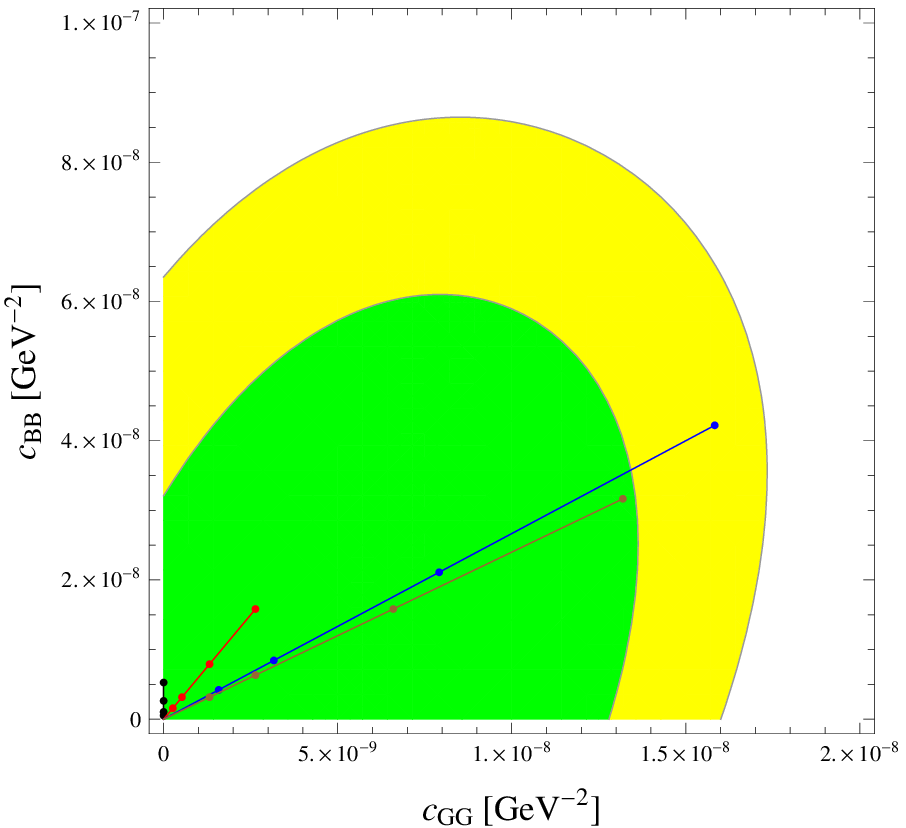}
\includegraphics[height=2.5in]{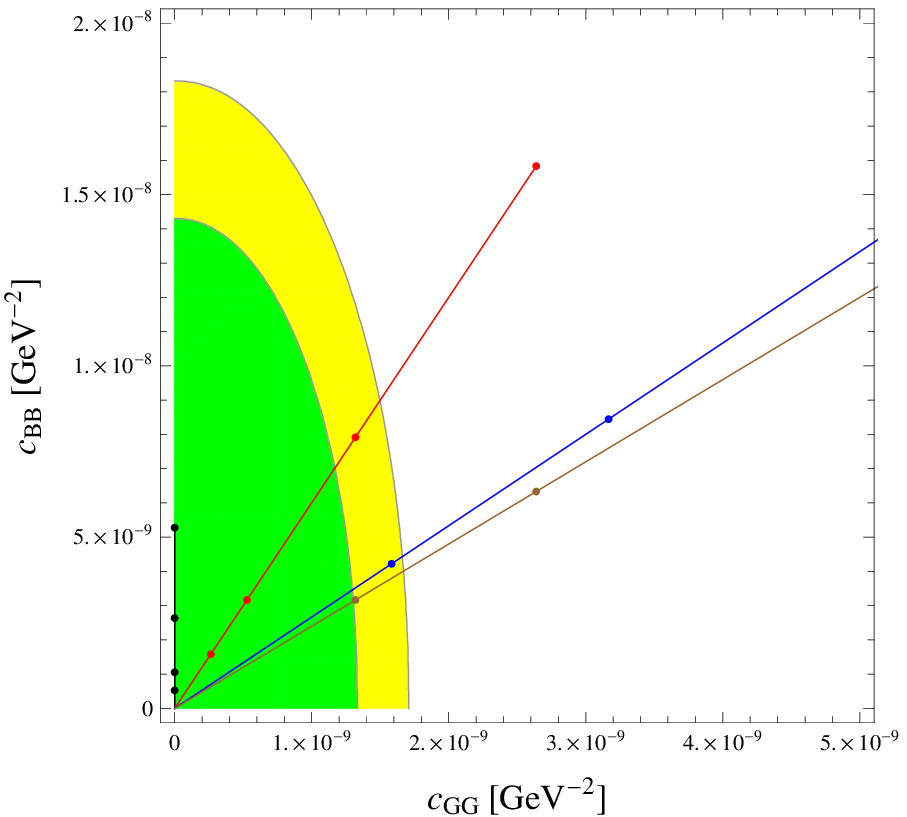}
\caption{Wilson coefficients $c_{GG}$ and $c_{BB}$ for contributions from a new $SU(2)_L$ singlet scalar field with $Y=1$ and in the $SU(3)_C$ representation of $\bf 1$ (black), $\bf 3$ (red), $\bf 8$ (blue) and $\bf 6$ (brown) at one-loop level.  The left (right) plot shows the current LHC constraints (the future prospect) at the $1\sigma$ (green) and $2\sigma$ (yellow) level.  The marks on each line correspond to $\lambda/M^2 = 10^{-6}$~GeV$^{-2}$, $2\times10^{-6}$~GeV$^{-2}$, $5\times10^{-6}$~GeV$^{-2}$ and $10^{-5}$~GeV$^{-2}$ from the origin outward.}
\label{fig:cGGcBB}
\end{figure}

The models in Groups~II and III of Table~\ref{tab:neutrino} have an $SU(2)_L$ singlet scalar charged under $SU(3)_C$ and $U(1)_Y$.  In this case, the electroweak oblique corrections vanish identically, and the dominant constraints come from the gluon fusion production of the SM Higgs boson and its digluon and diphoton decays.  That is, the theory parameter $\lambda/M^2$ is restricted by the allowed Wilson coefficients $c_{GG}$ and $c_{BB}$ in these models (mostly by the former).  The other Wilson coefficients $c_H$ (and $c_{2B}$) are nonzero but much less important.  Fig.~\ref{fig:cGGcBB} shows current and future constraints on $c_{GG}$ and $c_{BB}$ and the trajectories of their values from the contribution of an $SU(2)_L$ singlet scalar field by varying $\lambda/M^2$ up to $10^{-5}$ GeV$^{-2}$.  The new scalar field is assumed to carry hypercharge $Y = 1$ and transform under $SU(3)_C$ as {\bf 1} (black curve), {\bf 3} (red curve), {\bf 8} (blue), and {\bf 6} (brown).

For nontrivial representations of $SU(2)_L$ in the other groups of Table~\ref{tab:neutrino}, they involve the additional coupling $\lambda'$.  In this case, the $T$ parameter, which scales like $(\lambda'/M)^2$, plays an important role in restricting the parameter space.  Therefore, the bounds are usually for $\lambda'/M$ instead of $\lambda'/M^2$ from the $S$ parameter.

\section{Summary \label{sec:sum}}

Although it is widely believed that the standard model (SM) is at best a good effective theory at low energies, the fact that the observed 125-GeV Higgs boson has properties very close to that in the SM suggests that the new physics scale is high and the new degrees of freedom are likely to be in the decoupling limit.  Therefore, it is useful to work out an effective field theory (EFT) in terms of operators up to dimension 6 and composed of only the SM fields.

In this paper, we have analysed the EFT of the SM Higgs field for a wide class of weakly coupled renormalisable new physics models extended by one type of scalar fields and respecting CP symmetry, concentrating on the dimension-6 operators that have corrections to the electroweak oblique parameters and current Higgs observables.  We have shown that for the new scalar field of specific representations ($SU(2)_L$ singlet, doublet, and triplet), there are ``accidental'' interactions between the scalar and the SM Higgs fields that lead to dimension-6 operators at both tree and one-loop level.  For the scalar field of a general representation under the SM gauge groups, we have pointed out that there are only two generic quartic interactions that will lead to dimension-6 operators only at one-loop level.  We work out the Wilson coefficients associated with these operators for the general case in terms of the new physics parameters.

Using the existing LEP oblique parameter measurements and LHC Higgs data, we study the current constraints on the parameters of several benchmark models.  The same is also done for the projected results expected in the future experiment.  Although indirect, comparing the higher dimensional operators in the effective field theory with precision measurements is always a useful probe and complementary to the direct search method.

\section*{Acknowledgments}

RH is grateful to Kaladi Babu, Joan Elias-Mir{\'o} and Yongchao Zhang for useful discussions, and to Ligong Bian for sharing the code.  This research was supported in part by the Ministry of Science and Technology of Taiwan under Grant No.~MOST-100-2628-M-008-003-MY4 and the World Premier International Research Center Initiative, Ministry of Education, Culture, Sports, Science and Technology, Japan.


\end{document}